\documentclass[12pt,executivepaper]{article}
\usepackage{amsfonts}

\input{tcilatex}

\begin{document}

\begin{center}
{\Large Exactly Solvable Potentials by }$\mathcal{SO}\left( 2,2\right) $ 
{\Large Dynamical Algebra}

\bigskip

S.-A. Yahiaoui, M. Bentaiba\footnote{%
Corresponding author:
\par
E-mail address : bentaiba@hotmail.com
\par
\ \ \ \ \ \ \ \ \ \ \ \ \ \ \ \ \ \ \ \ \ \ \ bentaiba1@caramail.com
\par
\ \ \ \ \ \ \ \ \ \ \ \ \ \ \ \ \ \ \ \ \ \ \ bentaiba@mail.univ-blida.dz}

LPTHIRM, D\'{e}partement de Physique, Facult\'{e} des Sciences,

Universit\'{e} Saad DAHLAB de Blida, Algeria.

\bigskip

\textbf{Abstract}
\end{center}

The differential realization of the potential group $\mathcal{SO}\left(
2,2\right) $ is used. The spectrum-generating algebra for a kind of exactly
solvable potentials endowed with position-dependent mass is constructed.

\bigskip


$\ \ \ \ \ \ \ \ \ \ \ \ \ \ \ \ $Spectrum-generating algebras.


\bigskip

\section{Introduction}

Exact solutions for some quantum mechanical systems endowed with
position-dependent effective mass have attracted, in recent years, much
attention on behalf of physicists [1-10]. Effective mass Schr\"{o}dinger
equation was introduced by BenDaniel and Duke [1] in order to explain the
behavior of electrons in semiconductors. It have also many applications in
the fields of material sciences and condensed matter physics such as quantum
well and quantum dots [11], $^{3}H-$clusters [12], quantum liquids [13],
graded alloys [14], and heterostructures [15], etc.

In theoretical physics, various variety of methods and approaches are
employed for generating exactly solvable one-dimensional potentials such as
the supersymmetric quantum mechanical method [16] and group theory through
the Lie algebra approach [17-20]. Both of these two approaches give (almost)
identical bound-state energy eigenvalues. Supersymmetric quantum mechanics
have been recognized as the reformulation of the factorization method; i.e.
\ the generalization of the creation and annihilation operators given two
isospectral Hamiltonians, while Lie algebras provides us with an important
aspect from algebraic techniques which are used to construct the
Hamiltonian(s) from the Casimir operator(s) related to the group-algebraic
structures.

A different notion about spectrum-generating algebra (SGA) techniques, not
totally independent of preview concepts [22], was introduced by Cordero et
al.[23-26], and are based on the use of a realization of the infinitesimal
generators $\mathcal{J}_{i}$ $\left( i=0,+,-\right) $ in the non-compact
group $\mathcal{SO}\left( 2,1\right) $. Following their definition, the
generators of SGA can be used to replace the canonical variables in the Schr%
\"{o}dinger equation in such a way that the bound-state energy spectrum is
connected to the spectrum of a compact generators of the group [26]. A
convenient way to construct a SGA for physical systems is by introducing a
set of boson creation and annihilation operators [17], these later can be
recast into a set of generators of the underlying group. In terms of these
settings, the non-compact algebra appears either as a spectrum-generating
algebra or as a construction identified as potential algebra.

In the present article, we provide an extension of the algebraic treatments
within the context of the non-compact group $\mathcal{SO}\left( 2,2\right) $
using SGA to the description of the bound-state problems by connecting the
Hamiltonian of the supersymmetric quantum mechanics to that of the powerful
machinery of the group theory. It seems that considering the larger $%
\mathcal{SO}(2,2)$ potential algebra and applying the SGA scheme to them
could be instructive to recover a wide kind of exactly solvable (confluent)
Natanzon potentials defined by Natanzon [27]. With this, we generate and
"rediscover" all potentials generated with a similar differential
realization of the $\mathfrak{su}(1,1)\simeq \mathfrak{so}(2,1)$ algebra
[7,21,23,24]. It is well-known that the general families of the exactly
solvable potentials can be categorized in two different families : the
Natanzon and Natanzon confluent potentials. Each potentials in the same
family can be mapped into each others by applying point canonical
transformation (PCT) [9,10]. However, the link between potentials belonging
to the two different families is concerned with transformations larger than
PCT. The required transformations are identified as being integral
transformations [17,28].

The organization of the present article is as follows. In section 2 we
introduce a specific differential realization of the $\mathfrak{so}(2,2)$
algebra inspired by the potential-group method and derived from the
Hamiltonians which are expressed in terms of the Casimir operators of the
same algebra. We present and define also the general formulation of the
Hamiltonians endowed with position-dependent effective mass in the framework
of supersymmetric quantum mechanics. The main results are contained in
section 3, where the general expression of the effective potential is
deduced once the Hamiltonians of the previous sections are combined. In
section 4, we summarize our results by choosing the appropriate values for
the parameters $b$ and $q$ thus leading to generate quantum mechanical
effective potentials. Finally, the last section is devoted to some remarks
and a conclusion.

\section{Differential realization of the $\mathfrak{so}(2,2)$}

According to L\'{e}vai et al.[20], we start by giving a differential
realization of the $\mathfrak{so}(n,m)$ Lie algebras with six-generators
containing altogether ten functions. To be more precise, these differential
realizations are related to the algebras of $\mathfrak{so}(2,2)$, $\mathfrak{%
so}(4)$ and $\mathfrak{so}(3)\oplus \mathfrak{so}(2,1)$, allowing a more
general form of the generators divided into two sets%
\begin{eqnarray}
\mathcal{J}_{\pm } &=&\func{e}^{\pm i\phi }\left[ \pm h_{1}\left( x\right)
\partial _{x}\pm g_{1}\left( x\right) +f_{1}\left( x\right) \mathcal{J}%
_{0}+c_{1}\left( x\right) +k_{1}\left( x\right) \mathcal{L}_{0}\right] , 
\TCItag{1.a} \\
\mathcal{J}_{0} &=&-i\partial _{\phi },  \TCItag{1.b} \\
\mathcal{L}_{\pm } &=&\func{e}^{\pm i\chi }\left[ \pm h_{2}\left( x\right)
\partial _{x}\pm g_{2}\left( x\right) +f_{2}\left( x\right) \mathcal{L}%
_{0}+c_{2}\left( x\right) +k_{2}\left( x\right) \mathcal{J}_{0}\right] , 
\TCItag{1.c} \\
\mathcal{L}_{0} &=&-i\partial _{\chi },  \TCItag{1.d}
\end{eqnarray}%
where we have used the abbreviation $\partial _{\Sigma }=\frac{d}{d\Sigma }$%
, with $\Sigma =x,\phi ,\chi $. Then the commutation relations are given by%
\begin{eqnarray}
\left[ \mathcal{J}_{0},\mathcal{J}_{\pm }\right]  &=&\pm \mathcal{J}_{\pm
}\quad ;\quad \left[ \mathcal{J}_{+},\mathcal{J}_{-}\right] =-2a\mathcal{J}%
_{0},  \TCItag{2.a} \\
\left[ \mathcal{L}_{0},\mathcal{L}_{\pm }\right]  &=&\pm \mathcal{L}_{\pm
}\quad ;\quad \left[ \mathcal{L}_{+},\mathcal{L}_{-}\right] =-2b\mathcal{L}%
_{0},  \TCItag{2.b}
\end{eqnarray}%
\begin{equation}
\left[ \mathcal{J}_{i},\mathcal{L}_{j}\right] =0,\quad \left(
i,j=0,+,-\right) .  \tag{2.c}
\end{equation}%
where $a,b=\pm 1$. For $a=b=1$, we get $\mathfrak{so}(2,2)\simeq \mathfrak{so%
}(2,1)\oplus \mathfrak{so}(2,1)$ algebra, while for $a=-b=1$ and $a=b=-1$ we
obtain, respectively, the $\mathfrak{so}(3)\oplus \mathfrak{so}(2,1)$ and $%
\mathfrak{so}(4)\simeq \mathfrak{so}(3)\oplus \mathfrak{so}(3)$ algebras.
The above equations lead to a system of first-order equations for the
functions appearing in (1.a) and (1.c) [20]%
\begin{eqnarray}
k_{2}^{2}-h_{2}k_{2}^{\prime } &=&0;\quad h_{2}f_{2}^{\prime
}-f_{2}k_{2}=0;\quad k_{2}^{2}-f_{2}^{2}=b,  \TCItag{3.a} \\
h_{1} &=&Ah_{2};\quad f_{1}=Ak_{2};\quad k_{1}=Af_{2};\quad g_{1}=Ag_{2}, 
\TCItag{3.b} \\
c_{1} &=&c_{2}=0,  \TCItag{3.c}
\end{eqnarray}%
with $A^{2}ab=1$ and $b^{2}=1$. Here, we assume that $h_{i}\left( x\right)
\neq 0,$ $k_{i}\left( x\right) \neq 0$ and $f_{i}\left( x\right) \neq 0$,
with $i=1,2$. It is obvious, from (3.a), that the choice of the function $%
h_{2}\left( x\right) $ determines completely the shape of the functions $%
f_{2}\left( x\right) $ and $k_{2}\left( x\right) $.

An interesting way to solve the differential equations (3.a) is by applying
a variable transformation $x\rightarrow y\equiv y\left( x\right) $ which
changes the functions $h_{2}\left( x\right) $, $f_{2}\left( x\right) $ and $%
k_{2}\left( x\right) $ into the following forms [21]%
\begin{eqnarray}
h_{2}\left( x\right)  &\rightarrow &\widetilde{h}_{2}\left( x\right)
=h_{2}\left( x\right) \frac{dy\left( x\right) }{dx},  \TCItag{4.a} \\
k_{2}\left( x\right)  &\rightarrow &\widetilde{k}_{2}\left( x\right)
=k_{2}\left( x\right) ,  \TCItag{4.b} \\
f_{2}\left( x\right)  &\rightarrow &\widetilde{f}_{2}\left( x\right)
=f_{2}\left( x\right) ,  \TCItag{4.c} \\
g_{2}\left( x\right)  &\rightarrow &\widetilde{g}_{2}\left( x\right)
=g_{2}\left( x\right) ,  \TCItag{4.d} \\
c_{2}\left( x\right)  &\rightarrow &\widetilde{c}_{2}\left( x\right)
=c_{2}\left( x\right) .  \TCItag{4.e}
\end{eqnarray}

It is clear that the operators in (1.a) and (1.c) are maintained invariants
once the transformations (4) are applied. Now assuming that $h_{2}\left(
x\right) \frac{dy\left( x\right) }{dx}=\sqrt{b}y\left( x\right) $, the first
two differential equations in (3.a) become%
\begin{eqnarray}
\sqrt{b}y\left( x\right) f_{2}^{\prime }\left( x\right) -k_{2}\left(
x\right) f_{2}\left( x\right)  &=&0,  \TCItag{5.a} \\
k_{2}^{2}\left( x\right) -\sqrt{b}y\left( x\right) k_{2}^{\prime }\left(
x\right)  &=&0,  \TCItag{5.b}
\end{eqnarray}%
where their solutions are given, respectively, by%
\begin{equation}
k_{2}\left( x\right) =\sqrt{b}\frac{1+qy^{2}\left( x\right) /\sqrt{b}}{%
1-qy^{2}\left( x\right) /\sqrt{b}},\qquad f_{2}\left( x\right) =\frac{\delta
y\left( x\right) }{1-qy^{2}\left( x\right) /\sqrt{b}},  \tag{6}
\end{equation}%
where $q$ and $\delta $ are constants of integration.

Since the $\mathfrak{so}(2,2)$ algebra is of rank $2$, then it admits two
Casimir operators given by%
\begin{eqnarray}
\mathcal{C}_{\mathfrak{so}(2,2)}^{\left( \mathcal{J},\mathcal{L}\right) }
&=&2\mathcal{C}_{\mathfrak{so}(2,1)}^{\left( \mathcal{J}\right) }+2\mathcal{C%
}_{\mathfrak{so}(2,1)}^{\left( \mathcal{L}\right) },  \TCItag{7.a} \\
\widetilde{\mathcal{C}}_{\mathfrak{so}(2,2)}^{\left( \mathcal{J},\mathcal{L}%
\right) } &=&2\mathcal{C}_{\mathfrak{so}(2,1)}^{\left( \mathcal{J}\right) }-2%
\mathcal{C}_{\mathfrak{so}(2,1)}^{\left( \mathcal{L}\right) },  \TCItag{7.b}
\end{eqnarray}

A basis $\left\vert \lambda ,\nu ,\nu ^{\prime }\right\rangle $ for the
representation $\mathcal{SO}\left( 2,2\right) $ is characterized by%
\begin{eqnarray}
\mathcal{C}_{\mathfrak{so}(2,2)}^{\left( \mathcal{J},\mathcal{L}\right)
}\left\vert \lambda ,\nu ,\nu ^{\prime }\right\rangle  &=&\lambda \left(
\lambda +2\right) \left\vert \lambda ,\nu ,\nu ^{\prime }\right\rangle , 
\TCItag{8.a} \\
\mathcal{J}_{0}\left\vert \lambda ,\nu ,\nu ^{\prime }\right\rangle  &=&\nu
\left\vert \lambda ,\nu ,\nu ^{\prime }\right\rangle ,  \TCItag{8.b} \\
\mathcal{L}_{0}\left\vert \lambda ,\nu ,\nu ^{\prime }\right\rangle  &=&\nu
^{\prime }\left\vert \lambda ,\nu ,\nu ^{\prime }\right\rangle , 
\TCItag{8.c}
\end{eqnarray}%
and the basis $\left\vert \lambda ,\nu ,\nu ^{\prime }\right\rangle $,
namely eigenfunctions, can be written explicitly as%
\begin{equation}
\psi _{\nu ,\nu ^{\prime }}\left( x\right) =\exp \left[ i\left( \nu \phi
+\nu ^{\prime }\chi \right) \right] \mathcal{R}\left( x\right) ,  \tag{9}
\end{equation}

It turns out from Eqs.(7-9) that the eigenvalues of the first Casimir
operator are $\lambda \left( \lambda +2\right) $ where $\lambda $ is
connected with the eigenvalues $j\left( j+1\right) $ of the Casimir operator
of $\mathfrak{so}(2,1)$ through the relation $\lambda =2j$, while the second
Casimir operator (7.b) has always zero eigenvalues [20]. Here $\nu $ and $%
\nu ^{\prime }$ are the eigenvalues of the compact generators $\mathcal{J}%
_{0}$ and $\mathcal{L}_{0}$, respectively. The eigenfunction $\mathcal{R}%
\left( x\right) $ is the physical wavefunction depending only in $x$, while $%
\phi $ and $\chi $ are auxiliary variables [19,20]. In terms of this
realization, the Casimir operator (7.a) has the form%
\begin{eqnarray}
\mathcal{C}_{\mathfrak{so}(2,2)}^{\left( \mathcal{J},\mathcal{L}\right) }
&=&2\left[ -a\mathcal{J}_{+}\mathcal{J}_{-}+\mathcal{J}_{0}^{2}\mathcal{-J}%
_{0}-b\mathcal{L}_{+}\mathcal{L}_{-}+\mathcal{L}_{0}^{2}\mathcal{-L}_{0}%
\right]   \nonumber \\
&=&4bh_{2}^{2}\partial _{x}^{2}+4bh_{2}\left( h_{2}^{\prime
}+2g_{2}-k_{2}\right) \partial _{x}+4b\left( h_{2}g_{2}^{\prime
}+g_{2}^{2}-k_{2}g_{2}\right)   \nonumber \\
&&+2\left[ 1-b\left( k_{2}^{2}+f_{2}^{2}\right) \right] \left( \mathcal{J}%
_{0}^{2}\mathcal{+L}_{0}^{2}\right) -8bf_{2}k_{2}\mathcal{J}_{0}\mathcal{L}%
_{0}.  \TCItag{10}
\end{eqnarray}

Now, inserting (9) into (10) taking into account (6), we get%
\[
\mathcal{C}_{\mathfrak{so}(2,2)}^{\left( \mathcal{J},\mathcal{L}\right) }=4%
\frac{y^{2}}{y^{\prime 2}}\partial _{x}^{2}+4\frac{y}{y^{\prime }}\left[ 2%
\frac{g_{2}}{\sqrt{b}}-\frac{yy^{\prime \prime }}{y^{\prime 2}}-\frac{%
2qy^{2}/\sqrt{b}}{1-qy^{2}/\sqrt{b}}\right] \partial _{x}+4\left[ \frac{y}{%
y^{\prime }}\frac{g_{2}^{\prime }}{\sqrt{b}}\right. 
\]%
\begin{eqnarray}
&&\left. +\left( \frac{g_{2}}{\sqrt{b}}\right) ^{2}-\frac{1+qy^{2}/\sqrt{b}}{%
1-qy^{2}/\sqrt{b}}\frac{g_{2}}{\sqrt{b}}\right] +2\left( \nu ^{2}+\nu
^{\prime 2}\right) \left[ 1-\frac{b\delta ^{2}y}{\left( 1-qy^{2}/\sqrt{b}%
\right) ^{2}}\right.  \nonumber \\
&&\left. -\left( \frac{1+qy^{2}/\sqrt{b}}{1-qy^{2}/\sqrt{b}}\right) ^{2}%
\right] -8b\delta \sqrt{b}\nu \nu ^{\prime }y\frac{1+qy^{2}/\sqrt{b}}{\left(
1-qy^{2}/\sqrt{b}\right) ^{2}}.  \TCItag{11}
\end{eqnarray}

In the other hand, the general form of the Hamiltonians introduced by von
Roos [2] for the spatially varying mass $M\left( x\right) =m_{0}m\left(
x\right) $, where $m\left( x\right) $ is a dimensionless mass, read%
\begin{equation}
\mathcal{H}_{VR}=\frac{1}{4}\left[ m^{\eta }\left( x\right) pm^{\epsilon
}\left( x\right) pm^{\rho }\left( x\right) +m^{\rho }\left( x\right)
pm^{\epsilon }\left( x\right) pm^{\eta }\left( x\right) \right] +V\left(
x\right) ,  \tag{12}
\end{equation}%
where $m_{0}=1$ and the restriction on the parameters $\eta ,$ $\epsilon $
and $\rho $ checks the condition $\eta +\epsilon +\rho =-1$. Here $p\left(
\equiv -i\hbar \partial _{x}\right) $ is the momentum. In the natural units $%
\left( \hbar =c=1\right) $, the Hamiltonian $\mathcal{H}_{VR}$ becomes%
\begin{eqnarray}
\mathcal{H}_{VR} &=&-\frac{1}{4}\left[ \frac{2}{m}\partial _{x}^{2}-2\frac{%
m^{\prime }}{m}\partial _{x}-\left( 1+\epsilon \right) \frac{m^{\prime
\prime }}{m^{2}}+2\left\{ \eta \left( \eta +\epsilon +1\right) +\epsilon
+1\right\} \frac{m^{\prime 2}}{m^{3}}\right]   \nonumber \\
&&+V\left( x\right) .  \TCItag{13}
\end{eqnarray}

By introducing the eigenfunctions [7]%
\begin{equation}
\psi _{\sigma }\left( x\right) =2\sigma m\left( x\right) \frac{y^{2}\left(
x\right) }{y^{\prime 2}\left( x\right) }\mathcal{R}\left( x\right) , 
\tag{14}
\end{equation}%
where $\sigma \in 
%TCIMACRO{\U{211d} }%
%BeginExpansion
\mathbb{R}
%EndExpansion
$, letting now (13) acts on (14), and after some long and straightforward
algebra, we obtain%
\begin{eqnarray}
\mathcal{H}_{VR} &=&-\sigma \frac{y^{2}}{y^{\prime 2}}\partial _{x}^{2}-%
\frac{\sigma y}{y^{\prime }}\left[ 4+\frac{m^{\prime }y}{my^{\prime }}-\frac{%
4yy^{\prime \prime }}{y^{\prime 2}}\right] \partial _{x}+\frac{2\sigma y}{%
y^{\prime 2}}\left[ 3y^{\prime \prime }+\frac{yy^{\prime \prime \prime }}{%
y^{\prime }}\right.   \nonumber \\
&&\left. -\frac{3yy^{\prime \prime 2}}{y^{\prime 2}}\right] +\frac{\sigma
m^{\prime }y^{2}}{my^{\prime 2}}\left[ \frac{2\eta \left( yy^{\prime \prime
}-y^{\prime 2}\right) }{yy^{\prime }}-\left( 1+\eta \right) \left( \eta
+\epsilon \right) \frac{m^{\prime }}{m}\right]   \nonumber \\
&&-2\sigma +\frac{\sigma \left( \epsilon -1\right) m^{\prime \prime }}{2m}%
\frac{y^{2}}{y^{\prime 2}}+\frac{2\sigma my^{2}}{y^{\prime 2}}V\left(
x\right) .  \TCItag{15}
\end{eqnarray}

$\mathcal{C}_{\mathfrak{so}(2,2)}^{\left( \mathcal{J},\mathcal{L}\right) }$
and $\mathcal{H}_{VR}$ are related to the undetermined function $y\left(
x\right) $ which results, as we will see in the next section, on a simple
differential equation(s).

\section{Spectrum-generating algebra of $\mathfrak{so}(2,2)$}

The Schr\"{o}dinger equations can be solved once equating them to the
eigenvalues equation of the Casimir invariant operator of the $\mathfrak{so}%
(2,2)$ algebra [26]%
\begin{equation}
\left( \mathcal{H}_{VR}-E\right) \psi \left( x\right) =Z\left( x\right)
\left( \mathcal{C}_{\mathfrak{so}(2,2)}^{\left( \mathcal{J},\mathcal{L}%
\right) }-c\right) \psi \left( x\right) =0,  \tag{16}
\end{equation}%
where $Z\left( x\right) $ is some function to be determined, and $c$ is the
eigenvalue of the Casimir operator. Now substituting (11) and (15) into (16)
and comparing both sides we get%
\begin{eqnarray}
Z\left( x\right) &=&-\frac{\sigma }{4},  \TCItag{17.a} \\
\frac{g_{2}\left( x\right) }{\sqrt{b}} &=&\frac{2-qy^{2}\left( x\right) /%
\sqrt{b}}{1-qy^{2}\left( x\right) /\sqrt{b}}-\frac{3y\left( x\right)
y^{\prime \prime }\left( x\right) }{2y^{\prime 2}\left( x\right) }+\frac{%
m^{\prime }\left( x\right) y\left( x\right) }{2m\left( x\right) y^{\prime
}\left( x\right) }.  \TCItag{17.b}
\end{eqnarray}%
Inserting $g_{2}\left( x\right) $, $g_{2}^{\prime }\left( x\right) $ and $%
g_{2}^{2}\left( x\right) $ as defined in (17.b) into (16), taking into
consideration (11) and (15), we end up with%
\begin{eqnarray}
V_{\text{eff}}^{\left( q,b\right) }\left( x\right) -E_{\nu ,\nu ^{\prime
},c} &=&\left[ \frac{\nu ^{2}+\nu ^{\prime 2}}{4}\frac{4q/\sqrt{b}+b\delta
^{2}}{\left( 1-qy^{2}/\sqrt{b}\right) ^{2}}+b\delta \sqrt{b}\nu \nu ^{\prime
}\frac{1+qy^{2}/\sqrt{b}}{y\left( 1-qy^{2}/\sqrt{b}\right) ^{2}}\right. 
\nonumber \\
&&\left. +\frac{c}{8y^{2}}-\frac{q/\sqrt{b}}{2\left( 1-qy^{2}/\sqrt{b}%
\right) ^{2}}\right] \frac{y^{\prime 2}}{m}+\mathcal{V}_{\text{eff}}^{\left(
\eta ,\epsilon \right) }\left( x\right) ,  \TCItag{18}
\end{eqnarray}%
where by definition%
\begin{equation}
\mathcal{V}_{\text{eff}}^{\left( \eta ,\epsilon \right) }\left( x\right) =%
\frac{3}{8m}\frac{y^{\prime \prime 2}}{y^{\prime 2}}-\frac{1}{4m}\frac{%
y^{\prime \prime \prime }}{y^{\prime }}+\frac{m^{\prime 2}}{8m^{3}}\left[
\left( 1+2\eta \right) ^{2}+4\epsilon \left( 1+\eta \right) \right] -\frac{%
\epsilon m^{\prime \prime }}{4m^{2}},  \tag{19}
\end{equation}%
is called the effective potential related to the position-dependent mass and
depends on the parameters $\eta $ and $\epsilon $. In order to write
suitably (18), a new change of parameters $\left( \nu ,\nu ^{\prime }\right)
\rightarrow \left( r,t\right) $ is introduced%
\begin{equation}
\nu =\frac{1}{2}\left( \sqrt{r}+\sqrt{t}\right) \quad ;\quad \nu ^{\prime }=%
\frac{1}{2}\left( \sqrt{r}-\sqrt{t}\right) ,  \tag{20}
\end{equation}%
which brings (18) to%
\begin{eqnarray}
V_{\text{eff}}^{\left( q,b\right) }\left( x\right) -E_{r,t,c} &=&\frac{r}{8}%
\left[ \frac{4q/\sqrt{b}+b\delta ^{2}}{\left( 1-qy^{2}/\sqrt{b}\right) ^{2}}%
+2b\delta \sqrt{b}\frac{1+qy^{2}/\sqrt{b}}{y\left( 1-qy^{2}/\sqrt{b}\right)
^{2}}\right] \frac{y^{\prime 2}}{m}  \nonumber \\
&&+\frac{t}{8}\left[ \frac{4q/\sqrt{b}+b\delta ^{2}}{\left( 1-qy^{2}/\sqrt{b}%
\right) ^{2}}-2b\delta \sqrt{b}\frac{1+qy^{2}/\sqrt{b}}{y\left( 1-qy^{2}/%
\sqrt{b}\right) ^{2}}\right] \frac{y^{\prime 2}}{m}  \nonumber \\
&&+\frac{c}{8y^{2}}\frac{y^{\prime 2}}{m}-\frac{q/\sqrt{b}}{2\left( 1-qy^{2}/%
\sqrt{b}\right) ^{2}}\frac{y^{\prime 2}}{m}+\mathcal{V}_{\text{eff}}^{\left(
\eta ,\epsilon \right) }\left( x\right) .  \TCItag{21}
\end{eqnarray}

Without loss of generality, let us assume that the function $y\left(
x\right) $ is related to a certain \textit{generating }function, namely, $%
\mathfrak{S}\left( x\right) $ by%
\begin{equation}
\frac{y^{\prime 2}\left( x\right) }{m\left( x\right) }=\mathfrak{S}\left(
x\right) ,  \tag{22}
\end{equation}%
allowing us to rewrite (20) as%
\begin{eqnarray}
V_{\text{eff}}^{\left( q,b\right) }\left( x\right) -E_{r,t,c} &=&\frac{r}{8}%
F\left( x\right) \mathfrak{S}\left( x\right) +\frac{t}{8}G\left( x\right) 
\mathfrak{S}\left( x\right) +\frac{c}{8y^{2}\left( x\right) }\mathfrak{S}%
\left( x\right)   \nonumber \\
&&-\frac{q/\sqrt{b}}{2\left( 1-qy^{2}\left( x\right) /\sqrt{b}\right) ^{2}}%
\mathfrak{S}\left( x\right) +\mathcal{V}_{\text{eff}}^{\left( \eta ,\epsilon
\right) }\left( x\right) ,  \TCItag{23}
\end{eqnarray}%
where $F\left( x\right) $ and $G\left( x\right) $ are defined by%
\begin{eqnarray}
F\left( x\right)  &=&\frac{4q/\sqrt{b}+b\delta ^{2}}{\left( 1-qy^{2}\left(
x\right) /\sqrt{b}\right) ^{2}}+2b\delta \sqrt{b}\frac{1+qy^{2}\left(
x\right) /\sqrt{b}}{y\left( x\right) \left( 1-qy^{2}\left( x\right) /\sqrt{b}%
\right) ^{2}},  \TCItag{24.a} \\
G\left( x\right)  &=&\frac{4q/\sqrt{b}+b\delta ^{2}}{\left( 1-qy^{2}\left(
x\right) /\sqrt{b}\right) ^{2}}-2b\delta \sqrt{b}\frac{1+qy^{2}\left(
x\right) /\sqrt{b}}{y\left( x\right) \left( 1-qy^{2}\left( x\right) /\sqrt{b}%
\right) ^{2}}.  \TCItag{24.b}
\end{eqnarray}

Equation (23) is our main result, it is the key formula for generalized
potentials. Indeed it opens up ways to recover a wide kind of exactly
solvable potentials defined by Natanzon [27]. Observing that the energy $%
E_{r,t,c}$ on the left-hand side of (23) represents a constant term and,
therefore, must be equated to a certain constant term in the right-hand side
which has led to a simple differential equation for $y\left( x\right) $. On
the other hand, we can assume that the energy term $E_{r,t,c}$ can be
expressed in terms of the coefficients $r$, $t$ and $c$ which required that
both functions $y\left( x\right) $ and $\mathfrak{S}\left( x\right) $ be
independent of $E_{r,t,c}$.

On these settings, let us perform a formal derivative of (23) with respect
of $E_{r,t,c}$ and which henceforth will be quoted $E$, which leads to the
expression of the generating function $\mathfrak{S}\left( x\right) $ given by%
\begin{equation}
\mathfrak{S}\left( x\right) =\frac{-1}{\frac{F\left( x\right) }{8}\partial
_{E}r+\frac{G\left( x\right) }{8}\partial _{E}t+\frac{1}{8y^{2}\left(
x\right) }\partial _{E}c}.  \tag{25}
\end{equation}

Let the generating function $\mathfrak{S}\left( x\right) $ be positive, we
assume that the derivatives of the coefficients $r$, $t$ and $c$ with
respect of $E$ in (25) are constant, which requires that the coefficients
are linear with respect to $E$ [26]. In terms of these settings, the
coefficients become%
\begin{eqnarray}
\partial _{E}r &=&-r_{0}\quad \Longrightarrow \quad r\left( E\right)
=-r_{0}E+a_{r},  \TCItag{26.a} \\
\partial _{E}t &=&-t_{0}\quad \Longrightarrow \quad t\left( E\right)
=-t_{0}E+a_{t},  \TCItag{26.b} \\
\partial _{E}c &=&-c_{0}\quad \Longrightarrow \quad c\left( E\right)
=-c_{0}E+a_{c},  \TCItag{26.c}
\end{eqnarray}%
where $r_{0}$, $t_{0}$, $c_{0}$, $a_{r}$, $a_{t}$ and $a_{c}$ are six real
parameters. A straightforward algebraic manipulation permits to recast the
generating function $\mathfrak{S}\left( x\right) $ through%
\begin{eqnarray}
\mathfrak{S}\left( x\right)  &\equiv &\frac{y^{\prime 2}\left( x\right) }{%
m\left( x\right) }  \nonumber \\
&=&\frac{8y^{2}\left( x\right) }{r_{0}F\left( x\right) y^{2}\left( x\right)
+t_{0}G\left( x\right) y^{2}\left( x\right) +c_{0}}.  \TCItag{27}
\end{eqnarray}

One can see that the first three parameters $\left( r_{0},t_{0},c_{0}\right) 
$ govern completely the behavior of the function $y\left( x\right) $, while,
as we will see later, the last three parameters $\left(
a_{r},a_{t},a_{c}\right) $ determine the shape of the energy eigenvalues.

\section{Construction of exactly solvable potentials via SGA}

Here we illustrate the procedure by which a wide kind of exactly solvable
potentials belonging to the Natanzon and Natanzon confluent potentials can
be recovered using the SGA scheme. All these potentials fall into two cases
(classes) and are referred to (i) the appropriate-parameter choices of $%
\left( q,b\right) $ and $\left( r_{0},t_{0},c_{0}\right) $, and (ii) the
function $y\left( x\right) $. These parameter sets can be determined
arbitrary, while $y\left( x\right) $ can be obtained, after integration,
from the restriction (27). Among those choices, it is found that the
Natanzon confluent potentials [27] are generated when the constraints $q=0$
and $b=1$ are fulfilled, while the Natanzon potentials can be constructed
from $q=1$ and $b=1$.

\subsection{Case A : $q=0$, $b=1$}

In this subsection, the spectrum-generating algebra are explained in full
details. The functions $F\left( x\right) $ and $G\left( x\right) $ are given
by%
\begin{equation}
F\left( x\right) =\frac{\delta ^{2}y\left( x\right) +2\delta }{y\left(
x\right) }\quad ;\quad G\left( x\right) =\frac{\delta ^{2}y\left( x\right)
-2\delta }{y\left( x\right) },  \tag{28}
\end{equation}%
and thus, from (27), the generating function $\mathfrak{S}\left( x\right) $
becomes%
\begin{eqnarray}
\mathfrak{S}\left( x\right) &=&\frac{y^{\prime 2}\left( x\right) }{m\left(
x\right) }  \nonumber \\
&=&\frac{8y^{2}\left( x\right) }{c_{0}+2\delta \left( r_{0}-t_{0}\right)
y\left( x\right) +\delta ^{2}\left( r_{0}+t_{0}\right) y\left( x\right) }. 
\TCItag{29}
\end{eqnarray}

The Morse, Harmonic oscillator and Coulomb-like potentials can be deduced
from the generating function (29) and we will name them : exponential,
quadratic and linear solutions, respectively [7,21,27].

\subsubsection{Exponential solution : Morse-like potential}

This solution corresponds to the appropriate-parameter choices $c_{0}\neq 0$
and $r_{0}-t_{0}=r_{0}+t_{0}=0$, implying that $r_{0}=t_{0}=0$.
Consequently, the generating function $\mathfrak{S}\left( x\right) $ is then
determined from the differential equation $c_{0}y^{\prime 2}\left( x\right)
-8m\left( x\right) y^{2}\left( x\right) =0$, where their solutions take the
form%
\begin{equation}
y\left( x\right) =\exp \left[ -\alpha \mu \left( x\right) \right] ,  \tag{30}
\end{equation}%
where $\alpha =2\sqrt{\frac{2}{c_{0}}}$ and the auxiliary function $\mu
\left( x\right) =\int^{x}du\sqrt{m\left( u\right) }$ is defined, due to $%
m\left( x\right) $, as dimensionless mass integral. Substituting (26) and
(30) into (23), taking into consideration the restrictions concerning $r_{0}$%
, $t_{0}$ and $c_{0}$ referred hereabove, we get%
\begin{eqnarray}
V_{\text{eff}}^{\left( 0,1\right) }\left( x\right) -E &=&\frac{\delta
^{2}\alpha ^{2}}{8}\left( a_{r}+a_{t}\right) \func{e}^{-2\alpha \mu \left(
x\right) }+\frac{\delta \alpha ^{2}}{4}\left( a_{r}-a_{t}\right) \func{e}%
^{-\alpha \mu \left( x\right) }  \nonumber \\
&&+\frac{\alpha ^{2}}{8}\left( a_{c}-c_{0}E+1\right) +\mathcal{U}_{\text{eff}%
}^{\left( \eta ,\epsilon \right) }\left( x\right) ,  \TCItag{31}
\end{eqnarray}%
where the effective potential $\mathcal{U}_{\text{eff}}^{\left( \eta
,\epsilon \right) }\left( x\right) \equiv \mathcal{V}_{\text{eff}}^{\left(
\eta ,\epsilon \right) }\left( x\right) -\frac{\alpha ^{2}}{8}$ reads as%
\begin{equation}
\mathcal{U}_{\text{eff}}^{\left( \eta ,\epsilon \right) }\left( x\right) =%
\left[ 2\eta ^{2}+2\eta \left( 1+\epsilon \right) +\frac{3\epsilon }{2}+%
\frac{7}{8}\right] \frac{\mu ^{\prime \prime 2}\left( x\right) }{\mu
^{\prime 4}\left( x\right) }-\frac{1+2\epsilon }{4}\frac{\mu ^{\prime \prime
\prime }\left( x\right) }{\mu ^{\prime 3}\left( x\right) }.  \tag{32}
\end{equation}

It is obvious that we recognize in (31) the Morse effective-like potential
solely if the $a_{r}\neq \pm a_{t}$ constraint holds.

The energy eigenvalues can be deduced from (31) by equating the energy term
in the left-hand side with the constant term in the right-hand side. Indeed,
taking into account $\alpha =2\sqrt{\frac{2}{c_{0}}}$ and comparing both
sides, we obtain%
\begin{equation}
-E=\frac{\alpha ^{2}}{8}\left( a_{c}-c_{0}E+1\right) ,  \tag{33}
\end{equation}%
leading to the identification $a_{c}=-1$. In the other hand, it is
well-known that the eigenvalues of the Casimir operator associated to $%
\mathfrak{so}(2,2)$ algebra are $\lambda \left( \lambda +2\right) $, thus
the energy eigenvalues associated with the Morse potential can be deduced
using (26.c), i.e.%
\begin{eqnarray}
c\left( E\right)  &=&-c_{0}E+a_{c}  \nonumber \\
&=&-c_{0}E-1  \nonumber \\
&\equiv &\lambda \left( \lambda +2\right) ,  \TCItag{34}
\end{eqnarray}%
where finally the energy can be expressed in terms of $\alpha $ and $\lambda 
$ as%
\begin{equation}
E_{\lambda }=-\frac{\alpha ^{2}}{8}\left( 1+\lambda \right) ^{2}.  \tag{35}
\end{equation}

Since we are dealing with the Morse-type potential, let us choose a general
parameters $A$ and $B$ such that $\left[ \QDATOP{a_{r}}{a_{t}}\right]
=2B\left( \mp 2A\mp \alpha +\frac{2B}{\alpha ^{2}\delta ^{2}}\right) $ where 
$a_{r}\left( a_{t}\right) $ agrees with upper (lower) sign, respectively.
Therefore the Morse potential as defined in (31) can be written as%
\begin{equation}
V_{\text{M}}^{\left( 0,1\right) }\left( x\right) =B^{2}\exp \left[ -2\alpha
\mu \left( x\right) \right] -B\left( 2A+\alpha \right) \exp \left[ -\alpha
\mu \left( x\right) \right] .  \tag{36}
\end{equation}

\subsubsection{Quadratic solution : $3\mathcal{D}-$Harmonic oscillator-like
potential}

The quadratic solution depends on the following appropriate-parameter
choices $c_{0}=0$, $r_{0}=-t_{0}$ and $r_{0}\neq t_{0}$. The generating
function is reduced to the simple differential equation $\delta
r_{0}y^{\prime 2}\left( x\right) -2m\left( x\right) y\left( x\right) =0$
yielding the solution%
\begin{equation}
y\left( x\right) =\frac{\alpha }{4}\mu ^{2}\left( x\right) ,  \tag{37}
\end{equation}%
with $\alpha =\frac{2}{\delta r_{0}}$. Therefore, Eq.(26) becomes%
\begin{eqnarray}
V_{\text{eff}}^{\left( 0,1\right) }\left( x\right) -E &=&\frac{\delta
^{2}\alpha ^{2}}{32}\left( a_{r}+a_{t}\right) \mu ^{2}\left( x\right) +\frac{%
a_{c}+\frac{3}{4}}{2\mu ^{2}\left( x\right) }-\frac{\delta \alpha r_{0}E}{2}
\nonumber \\
&&+\left( a_{r}-a_{t}\right) \frac{\delta \alpha }{4}+\mathcal{V}_{\text{eff}%
}^{\left( \eta ,\epsilon \right) }\left( x\right) ,  \TCItag{38}
\end{eqnarray}%
where we recognize the $3\mathcal{D}-$harmonic oscillator-like potential.

The procedure to generate the energy eigenvalues corresponding to the $3%
\mathcal{D}-$Harmonic oscillator is exactly as before. Equating, in (38),
the energy term in the left-hand side with the constant term in the
right-hand side, taking into account the restrictions $\alpha =\frac{2}{%
\delta r_{0}}$, $c_{0}=0$ and $r_{0}=-t_{0}$, we obtain%
\begin{equation}
-\left( 1-\frac{r_{0}\delta \alpha }{2}\right) E=\left( a_{r}-a_{t}\right) 
\frac{\delta \alpha }{4}=0,  \tag{39}
\end{equation}%
which results to identify that $a_{r}=a_{t}$. Dealing with the $3\mathcal{D}-
$Harmonic oscillator potential, let us assume $a_{r}=a_{t}=1$ and $\delta
\alpha =2\omega $ implying that $r_{0}=-t_{0}=\omega ^{-1}$. By combining
(26.a) to (26.b), we can deduce the analytic expression of the energy%
\begin{equation}
E_{\nu ,\nu ^{\prime }}=\delta \alpha \nu \nu ^{\prime }\quad
\Longrightarrow \quad E_{n}=2\omega n.  \tag{40}
\end{equation}%
where $n=\nu \nu ^{\prime }$ is the \textit{reduced} quantum number for the $%
3\mathcal{D}-$Harmonic oscillator under the $\mathfrak{so}(2,2)$ algebra.
Knowing from (26.c) that $c\left( E\right) \equiv a_{c}=\lambda \left(
\lambda +2\right) $, the potential in (38) can be reduced to%
\begin{equation}
V_{\text{H.O}}^{\left( 0,1\right) }\left( x\right) =\frac{1}{4}\omega
^{2}\mu ^{2}\left( x\right) +\frac{3+4\lambda \left( \lambda +2\right) }{%
8\mu ^{2}\left( x\right) }.  \tag{41}
\end{equation}

\subsubsection{Linear solution : $3\mathcal{D}-$Coulomb-like potential}

The linear solution corresponds to the specific-parameter choices $c_{0}=0$, 
$r_{0}=t_{0}$ and $r_{0}\neq -t_{0}$. The generating function is reduced to
the following differential equation $\delta ^{2}r_{0}y^{\prime 2}\left(
x\right) -4m\left( x\right) =0$, from where the solution is given by $%
y\left( x\right) =\alpha \mu \left( x\right) $ with $\alpha =\frac{2}{\delta 
\sqrt{r_{0}}}$. The effective potential reads as%
\begin{equation}
V_{\text{eff}}^{\left( 0,1\right) }\left( x\right) -E=\frac{\delta \alpha
\left( a_{r}-a_{t}\right) }{2\mu \left( x\right) }+\frac{a_{c}}{8\mu
^{2}\left( x\right) }-\frac{r_{0}\delta ^{2}\alpha ^{2}E}{4}+\frac{\delta
^{2}\alpha ^{2}}{8}\left( a_{r}+a_{t}\right) +\mathcal{V}_{\text{eff}%
}^{\left( \eta ,\epsilon \right) }\left( x\right) ,  \tag{42}
\end{equation}%
where we recognize here the three-dimensional Coulomb-like potential.

The energy eigenvalues accompanying the potential (42) can be deduced once
equating the constant terms involving in (42), and after some algebraic
manipulations we obtain%
\begin{eqnarray}
a_{r} &=&-a_{t},  \TCItag{43.a} \\
c\left( E\right)  &\equiv &a_{c}=\lambda \left( \lambda +2\right) , 
\TCItag{43.b}
\end{eqnarray}%
and combining, once again, (26.a) to (26.b) we deduce the analytic
expression of the energy related to the $3\mathcal{D}-$Coulomb potential%
\begin{equation}
E_{\nu ,\nu ^{\prime }}=-\frac{\nu ^{2}+\nu ^{\prime 2}}{r_{0}},  \tag{44}
\end{equation}%
where $r_{0}=\frac{4}{\delta ^{2}\alpha ^{2}}$. Since we are dealing with
three-dimensional Coulomb potential, we will assume that $Ze^{2}=\delta
\alpha \nu \nu ^{\prime }$, then Eq.(44) becomes%
\begin{equation}
E_{\nu ,\nu ^{\prime }}=-\frac{Z^{2}e^{4}}{4}\left[ \frac{1}{\nu ^{2}}+\frac{%
1}{\nu ^{\prime 2}}\right] \quad \Longrightarrow \quad E_{\mathcal{N}}=-%
\left[ \frac{Ze^{2}}{2\mathcal{N}}\right] ^{2},  \tag{45}
\end{equation}%
with $\mathcal{N=}\frac{\nu \nu ^{\prime }}{\sqrt{\nu ^{2}+\nu ^{\prime 2}}}$
is the reduced quantum number for the $3\mathcal{D}-$Coulomb potential under 
$\mathfrak{so}(2,2)$ algebra. Knowing from (26.c) that $c\left( E\right)
\equiv a_{c}=\lambda \left( \lambda +2\right) $ and $a_{r}=-a_{t}=-Ze^{2}$
the potential in (42) can be written as%
\begin{equation}
V_{\text{Cb}}^{\left( 0,1\right) }\left( x\right) =-\frac{Ze^{2}}{\mu \left(
x\right) }+\frac{\lambda \left( \lambda +2\right) }{8\mu ^{2}\left( x\right) 
}.  \tag{46}
\end{equation}

\subsection{Case B : $q=1$, $b=1$}

In this case, the functions $F\left( x\right) $ and $G\left( x\right) $
become, from Eqs.(24)%
\begin{eqnarray}
F\left( x\right) &=&\frac{\left( 4+\delta ^{2}\right) y\left( x\right)
+2\delta \left( 1+y^{2}\left( x\right) \right) }{y\left( x\right) \left(
1-y^{2}\left( x\right) \right) ^{2}},  \TCItag{47.a} \\
G\left( x\right) &=&\frac{\left( 4+\delta ^{2}\right) y\left( x\right)
-2\delta \left( 1+y^{2}\left( x\right) \right) }{y\left( x\right) \left(
1-y^{2}\left( x\right) \right) ^{2}},  \TCItag{47.b}
\end{eqnarray}%
and the generating function $\mathfrak{S}\left( x\right) $ is given by%
\begin{eqnarray}
\mathfrak{S}\left( x\right) &\equiv &\frac{y^{\prime 2}\left( x\right) }{%
m\left( x\right) }  \nonumber \\
&=&\frac{8y^{2}\left( x\right) \left( 1-y^{2}\left( x\right) \right) ^{2}}{%
c_{0}\left( 1-y^{2}\left( x\right) \right) ^{4}+\left( r_{0}+t_{0}\right)
\left( 4+\delta ^{2}\right) y^{2}\left( x\right) +2\delta \left(
r_{0}-t_{0}\right) y\left( x\right) \left( 1-y^{2}\left( x\right) \right) }.
\TCItag{48}
\end{eqnarray}

Two particular cases arise and are associated to the appropriate-parameters
choices of $r_{0}$, $t_{0}$, and $c_{0}$. Thus a wide kind of exactly
solvable potentials, belonging to the Natanzon potentials, can be deduced
such as the\ generalized P\"{o}schl-Teller, P\"{o}schl-Teller, Scarf II,
Eckart, Hulth\`{e}n and Rosen-Morse-like potentials as well as their
trigonometric versions [7,21,27]. Here we present our results without giving
the details of our calculations which are straightforward and are exactly as
before.

\subsubsection{Case B-I : $r_{0}=t_{0}=0$, $c_{0}\neq 0$}

\paragraph{The generalized P\"{o}schl-Teller potential and its eigenvalues.}

Here the generating function (48) has as solution%
\begin{equation}
y\left( x\right) =\exp \left[ -\alpha \mu \left( x\right) \right] ,  \tag{49}
\end{equation}%
with $\alpha =2\sqrt{\frac{2}{c_{0}}}$. Consequently, the deduced potential
and its corresponding energy eigenvalues are given, respectively, by%
\begin{eqnarray}
V_{\text{GPT}}^{\left( 1,1\right) }\left( x\right) &=&\frac{\alpha ^{2}}{32}%
\left[ \left( 4+\delta ^{2}\right) \left( a_{r}+a_{t}\right) -4\right] \func{%
cosech}^{2}\alpha \mu \left( x\right)  \nonumber \\
&&+\frac{\delta \alpha ^{2}\left( a_{r}-a_{t}\right) }{8}\func{cosech}\alpha
\mu \left( x\right) \coth \alpha \mu \left( x\right) ,  \TCItag{50}
\end{eqnarray}%
\begin{equation}
E_{\lambda }=-\frac{\alpha ^{2}}{8}\left( 1+\lambda \right) ^{2}.  \tag{51}
\end{equation}

If the restriction $a_{r}\neq \pm a_{t}$ holds, then we recognize here the
generalized P\"{o}schl-Teller-like potential. The extension towards its
trigonometric version is possible once the substitution $\alpha \rightarrow
i\alpha $ is made leading to%
\begin{eqnarray}
V_{\text{GPT}}^{\left( 1,1\right) }\left( x\right)  &=&\frac{\alpha ^{2}}{32}%
\left[ \left( 4+\delta ^{2}\right) \left( a_{r}+a_{t}\right) -4\right] \csc
^{2}\alpha \mu \left( x\right)   \nonumber \\
&&+\frac{\delta \alpha ^{2}\left( a_{r}-a_{t}\right) }{8}\csc \alpha \mu
\left( x\right) \cot \alpha \mu \left( x\right) ,  \TCItag{52}
\end{eqnarray}%
\begin{equation}
E_{\lambda }=\frac{\alpha ^{2}}{8}\left( 1+\lambda \right) ^{2}.  \tag{53}
\end{equation}

\paragraph{The P\"{o}schl-Teller potential and its eigenvalues.}

In order to deduce the P\"{o}schl-Teller potential, let us multiply the
exponent of (49) by $2$, i.e. $y\left( x\right) =\exp \left[ -2\alpha \mu
\left( x\right) \right] $, that implies to redefine a new parameter $\alpha $
as $\alpha \rightarrow \alpha =\sqrt{\frac{2}{c_{0}}}$. The potential becomes%
\begin{eqnarray}
V_{\text{PT}}^{\left( 1,1\right) }\left( x\right) &=&-\frac{\alpha ^{2}}{32}%
\left[ \left( \delta +2\right) ^{2}a_{t}+\left( \delta -2\right) ^{2}a_{r}-4%
\right] \func{sech}^{2}\alpha \mu \left( x\right)  \nonumber \\
&&+\frac{\alpha ^{2}}{32}\left[ \left( \delta +2\right) ^{2}a_{r}+\left(
\delta -2\right) ^{2}a_{t}-4\right] \func{cosech}^{2}\alpha \mu \left(
x\right) ,  \TCItag{54}
\end{eqnarray}%
and its accompanying energy eigenvalues are given by%
\begin{equation}
E_{\lambda }=-\frac{\alpha ^{2}}{2}\left( 1+\lambda \right) ^{2}.  \tag{55}
\end{equation}

We have generated here the hyperbolic P\"{o}schl-Teller-like potential. Its
trigonometric version is obtained once the substitution $\alpha \rightarrow
i\alpha $ is carried out, given by%
\begin{eqnarray}
V_{\text{PT}}^{\left( 1,1\right) }\left( x\right)  &=&\frac{\alpha ^{2}}{32}%
\left[ \left( \delta +2\right) ^{2}a_{t}+\left( \delta -2\right) ^{2}a_{r}-4%
\right] \sec ^{2}\alpha \mu \left( x\right)   \nonumber \\
&&+\frac{\alpha ^{2}}{32}\left[ \left( \delta +2\right) ^{2}a_{r}+\left(
\delta -2\right) ^{2}a_{t}-4\right] \csc ^{2}\alpha \mu \left( x\right) , 
\TCItag{56}
\end{eqnarray}%
\begin{equation}
E_{\lambda }=\frac{\alpha ^{2}}{2}\left( 1+\lambda \right) ^{2}.  \tag{57}
\end{equation}

\paragraph{The Scarf II potential and its eigenvalues.}

The last potential obtainable in this category relates to the choice $%
y\left( x\right) =\exp \left[ -\alpha \mu \left( x\right) +\frac{i\pi }{2}%
\right] $. Due to the shape of $y\left( x\right) $, this potential is called 
\textit{isospectral} to the generalized P\"{o}schl-Teller potential because
they share the same energy eigenvalues. The potential and its corresponding
energy eigenvalues read, respectively%
\begin{eqnarray}
V_{\text{SC}}^{\left( 1,1\right) }\left( x\right) &=&-\frac{\alpha ^{2}}{32}%
\left[ \left( 4+\delta ^{2}\right) \left( a_{r}+a_{t}\right) -4\right] \func{%
sech}^{2}\alpha \mu \left( x\right)  \nonumber \\
&&+i\frac{\delta \alpha ^{2}\left( a_{r}-a_{t}\right) }{8}\func{sech}\alpha
\mu \left( x\right) \tanh \alpha \mu \left( x\right) ,  \TCItag{58}
\end{eqnarray}%
\begin{equation}
E_{\lambda }=-\frac{\alpha ^{2}}{8}\left( 1+\lambda \right) ^{2}.  \tag{59}
\end{equation}

We generate here the hyperbolic $\mathcal{PT}-$symmetric Scarf-like
potential. On can see that the energy eigenvalues (51) and (59) coincide,
thus it is then obvious that the generalized P\"{o}schl-Teller and the Scarf
II potentials are isospectral. The non$-\mathcal{PT}-$symmetric potential,
corresponding to the trigonometric case, is obtained once the substitution $%
\alpha \rightarrow i\alpha $ is made, given%
\begin{eqnarray}
V_{\text{SC}}^{\left( 1,1\right) }\left( x\right) &=&\frac{\alpha ^{2}}{32}%
\left[ \left( 4+\delta ^{2}\right) \left( a_{r}+a_{t}\right) -4\right] \sec
^{2}\alpha \mu \left( x\right)  \nonumber \\
&&+i\frac{\delta \alpha ^{2}\left( a_{r}-a_{t}\right) }{8}\sec \alpha \mu
\left( x\right) \tan \alpha \mu \left( x\right) ,  \TCItag{60}
\end{eqnarray}%
\begin{equation}
E_{\lambda }=\frac{\alpha ^{2}}{8}\left( 1+\lambda \right) ^{2}.  \tag{61}
\end{equation}

\subsubsection{Case B-II : $r_{0}=t_{0}\neq 0$, $c_{0}=0$}

Here, the generating function (48) is reduced to%
\begin{equation}
\mathfrak{S}\left( x\right) \equiv \frac{y^{\prime 2}\left( x\right) }{%
m\left( x\right) }=\frac{4p\left( 1-y^{2}\left( x\right) \right) ^{2}}{%
r_{0}\left( 4+\delta ^{2}\right) },  \tag{62}
\end{equation}%
where the class of solutions for the differential equation (62) dependent on
the arbitrary parameter $p=1,2,\ldots $ are given through%
\begin{equation}
y_{p}\left( x\right) =\tanh \frac{\alpha \mu \left( x\right) }{2p},  \tag{63}
\end{equation}%
with $\alpha =\frac{4p}{\sqrt{r_{0}\left( 4+\delta ^{2}\right) }}$.

\paragraph{The Eckart potential and its eigenvalues.}

For $p=1$ and taking into consideration the hyperbolic identity%
\begin{equation}
\frac{1+\tanh ^{2}a}{\tanh \frac{a}{2}}=2\coth a,  \tag{64}
\end{equation}%
one deduces the potential and its accompanying energy eigenvalues,
respectively%
\begin{equation}
V_{\text{Eck}}^{\left( 1,1,1\right) }\left( x\right) =-\frac{\delta \alpha
^{2}a_{t}}{4}\coth \alpha \mu \left( x\right) +\frac{\alpha ^{2}\lambda
\left( \lambda +2\right) }{8}\func{cosech}^{2}\alpha \mu \left( x\right) , 
\tag{65}
\end{equation}%
\begin{equation}
E_{\nu ,\nu ^{\prime }}=-\frac{\alpha ^{2}}{16}\left( 4+\delta ^{2}\right)
\left( \nu ^{2}+\nu ^{\prime 2}\right) .  \tag{66}
\end{equation}

We recognize here the Eckart-like potential. In the other hand its
trigonometric version is obtained once both substitutions $\alpha
\rightarrow i\alpha $ and $\delta \rightarrow -i\delta $ are made, given%
\begin{equation}
V_{\text{Eck}}^{\left( 1,1,1\right) }\left( x\right) =-\frac{\delta \alpha
^{2}a_{t}}{4}\cot \alpha \mu \left( x\right) +\frac{\alpha ^{2}\lambda
\left( \lambda +2\right) }{8}\csc ^{2}\alpha \mu \left( x\right) ,  \tag{67}
\end{equation}%
\begin{equation}
E_{\nu ,\nu ^{\prime }}=\frac{\alpha ^{2}}{16}\left( 4-\delta ^{2}\right)
\left( \nu ^{2}+\nu ^{\prime 2}\right) .  \tag{68}
\end{equation}

\paragraph{The Hulth\`{e}n potential and its eigenvalues.}

For $p=2$, we generate the following potential%
\begin{equation}
V_{\text{Hult}}^{\left( 1,1,2\right) }\left( x\right) =\frac{\alpha ^{2}}{4}%
\left[ \lambda \left( \lambda +2\right) +\frac{\delta \nu \nu ^{\prime }}{2}%
\right] \frac{\func{e}^{-\alpha \mu \left( x\right) }}{2\left( 1-\func{e}%
^{-\alpha \mu \left( x\right) }\right) }+\frac{\alpha ^{2}}{4}\lambda \left(
\lambda +2\right) \frac{\func{e}^{-2\alpha \mu \left( x\right) }}{2\left( 1-%
\func{e}^{-\alpha \mu \left( x\right) }\right) ^{2}},  \tag{69}
\end{equation}%
and its associated energy eigenvalues are given by%
\begin{equation}
E_{\nu ,\nu ^{\prime }}=-\frac{\alpha ^{2}}{16}\left[ \left( 4+\delta
^{2}\right) \left( \nu ^{2}+\nu ^{\prime 2}\right) +4\delta \nu \nu ^{\prime
}\right] .  \tag{70}
\end{equation}

The expressions (69) and (70) are known in the literature as being the Hulth%
\`{e}n potential and energy eigenvalues, respectively.

\paragraph{The Rosen-Morse potential and its eigenvalues.}

Now we are interested in the isospectral potential which is associated to
the Eckart potential with the coefficient $\frac{i\pi }{4}$. Consequently,
with $p=1$, the generating function becomes $y\left( x\right) =\tanh \left[ 
\frac{\alpha \mu \left( x\right) }{2}+\frac{i\pi }{4}\right] $. In this
respect, the use of the hyperbolic identity%
\begin{equation}
\tanh \left( a+ib\right) =\frac{\tanh a+i\tan b}{1+i\tanh a\tan b},  \tag{71}
\end{equation}%
led, after a long calculation, to deduce the following potential%
\begin{equation}
V_{\text{RM}}^{\left( 1,1,1\right) }\left( x\right) =\frac{\delta \alpha ^{2}%
}{8}\left( a_{r}-a_{t}\right) \tanh \alpha \mu \left( x\right) -\frac{\alpha
^{2}}{8}\lambda \left( \lambda +2\right) \func{sech}^{2}\alpha \mu \left(
x\right) ,  \tag{72}
\end{equation}%
where their accompanying energy eigenvalues are given%
\begin{equation}
E_{\nu ,\nu ^{\prime }}=-\frac{\alpha ^{2}}{16}\left( 4+\delta ^{2}\right)
\left( \nu ^{2}+\nu ^{\prime 2}\right) .  \tag{73}
\end{equation}%
The potential (72) and the energy eigenvalues (73) are those of the
Rosen-Morse type. Indeed one can see that the Eckart and the Rosen-Morse
potentials are isospectral, i.e. sharing the same energy eigenvalues.

\section{Conclusion}

In this article, we have discussed the general differential realization of
the potential group $\mathcal{SO}(2,2)$ and we have solved the Schr\"{o}%
dinger equation by identifying it to the eigenvalues equation of the Casimir
invariant operator(s) of this algebra. We have analyzed the role of the two $%
\mathfrak{so}(2,1)$ subalgebras and point out their importance by means of
spectrum-generating algebra techniques for a wide kind of potentials, for
which an analytic solution to the bound-state problem have been deduced;
i.e. all potentials defined by Natanzon [27]. We have constructed the case A
corresponding to $q=0$ and $b=1$ in full detail thus generating the Morse,
Harmonic oscillator and Coulomb-like potentials, while the case B with $q=1$
and $b=1$ generates the generalized P\"{o}schl-Teller, P\"{o}schl-Teller,
Scarf II, Eckart, Hulth\`{e}n and Rosen-Morse-like potentials as well as
their trigonometric versions. In terms of these settings, it becomes clear
that the former (case A) corresponds to the Natanzon confluent potentials
including the confluent hypergeometric functions as wavefunctions, while the
latter (case B) coincides with the Natanzon potentials and their
wavefunctions include hypergeometric functions.

In the light of all these, it has been shown that the exact solution of the
position-dependent effective mass Schr\"{o}dinger equation leads to a
general solution of the Natanzon potentials, which are independent of the
choice of the parameters $\eta $, $\epsilon $ and $\rho $.

\end{document}